# Synthesis, Structural and Electronic Properties of K$_4$Pu$^{VI}$O$_2$(CO$_3$)$_{3(cr)}$: Environmentally Relevant Plutonium Carbonate Complex

*Ivan Pidchenko,*[a,b] *Juliane März,*[b] *Myrtille O. J. Y. Hunault,*[c] *Sergei M. Butorin,*[d] *and Kristina O. Kvashnina*[a,b]

[a] Rossendorf Beamline at ESRF – The European Synchrotron, CS40220, 38043 Grenoble Cedex 9, France

[b] Helmholtz Zentrum Dresden-Rossendorf (HZDR), Institute of Resource Ecology, PO Box 510119, 01314, Dresden

[c] Synchrotron SOLEIL, L'Orme des Merisiers, Saint Aubin BP 48, 91192 Gif-sur-Yvette, France

[d] Molecular and Condensed Matter Physics, Department of Physics and Astronomy, Uppsala University, P.O. Box 516, Uppsala, Sweden

**Abstract:** Chemical properties of actinide materials are often pre-defined and described, based on the data available for isostructural species. This is the case for the potassium plutonyl (Pu$^{VI}$) carbonate - K$_4$Pu$^{VI}$O$_2$(CO$_3$)$_{3(cr)}$ - complex relevant for nuclear technology and environment, which crystallographic and thermodynamic properties are still lacking. We report here the synthesis and characterization of the K$_4$Pu$^{VI}$O$_2$(CO$_3$)$_{3(cr)}$ achieved by single-crystal X-ray diffraction analysis and high energy resolution fluorescence detected (HERFD) X-ray absorption near edge structure (XANES) at the Pu M$_4$ edge coupled with electronic structure calculations. Crystallographic properties of K$_4$Pu$^{VI}$O$_2$(CO$_3$)$_{3(cr)}$ are compared with isostructural uranium (U) and neptunium (Np) compounds. Actinyl (An$^{VI}$) axial bond lengths, [O–An$^{VI}$–O]$^{2+}$, are correlated between solid, K$_4$An$^{VI}$O$_2$(CO$_3$)$_{3(cr)}$, and aqueous, [An$^{VI}$O$_2$(CO$_3$)$_3$]$^{4-}_{(aq)}$ species for the U$^{VI}$–Np$^{VI}$–Pu$^{VI}$ series. The spectroscopic data are compared to the KPu$^{V}$O$_2$CO$_{3(cr)}$ and Pu$^{IV}$O$_{2(cr)}$ to follow the trend in the electronic structure of K$_4$Pu$^{VI}$O$_2$(CO$_3$)$_{3(cr)}$ regarding the oxidation state changes and local structural modifications around Pu atom.

Plutonium is the element of serious environmental concern due to the issues associated with the long-term storage of nuclear materials, including spent nuclear fuel.[1] Significant amounts of plutonium are left as a legacy after the fabrication and reprocessing of the nuclear materials and require adequate solutions for their securing and remediation of contaminated sites.[2] Whilst tetravalent plutonium in the form of colloidal Pu$^{IV}$O$_2$ is thought as the most environmentally relevant and mobile form of plutonium;[3,4] under certain conditions higher plutonium oxidation states, i.e. plutonyl, [Pu$^{VI}$O$_2$]$^{2+}$, may form in the groundwaters in oxic zones.[5] A large class of plutonium compounds is represented by environmentally relevant compounds e.g. hydroxide, fluorides, carbonates, systematically studied with other actinides in several countries and laboratories[6-9]. Actinides carbonates are assumed to greatly control speciation and geochemical mobility of actinides for neutral pH and alkaline environments by forming thermodynamically stable carbonate complexes: [An$^{VI}$O$_2$(CO$_3$)$_n$]$^{2n-2}$. The [An$^{VI}$O$_2$(CO$_3$)$_3$]$^{4-}$ dominates exclusively in waters with high carbonate content, [CO$_3^{2-}$] ≥ 10$^{-3}$ M, where solid complexes with composition M$_4$[An$^{VI}$O$_2$(CO$_3$)$_3$], M = Na, K, might potentially form.[10] The M$_4$[An$^{VI}$O$_2$(CO$_3$)$_3$] is an important class of compounds in the nuclear and reprocessing technology for the separation of the actinides along with An$^{VI}$ complexes with other ligands in the course of fundamental actinide science.[11-14]

Transuranic compounds remain less studied since most investigations are conducted on uranium due to the availability and relatively low radiotoxicity of uranium element. Hence the characterization of some environmentally relevant actinide species in higher valence states is lacking. To fill this gap, we performed the synthesis of K$_4$Pu$^{VI}$O$_2$(CO$_3$)$_{3(cr)}$ (further as **Pu$^{VI}$**) by slow evaporation of [Pu$^{VI}$O$_2$]$^{2+}$ in 1M K$_2$CO$_3$ solution (see Supporting Information (SI) for more details). Pu redox preparations were done using an *in situ* spectro-electrochemical setup starting from Pu$^{IV}$$_{aq}$ in 1M HClO$_4$ solution (see Figure S1a-c). Crystal structure refinement of **Pu$^{VI}$** was done by single-crystal X-ray diffraction (SC-XRD) and further characterized with high energy resolution fluorescence detected X-ray absorption near edge structure (HERFD-XANES) technique at the Pu M$_4$ edge.[15] For this a few green crystals of **Pu$^{VI}$** were placed into a double confinement sample holder designed for An M$_{4,5}$ edges measurements. Experimentally obtained HERFD-XANES data is analysed and supported by theoretical calculations based on the Anderson Impurity Model (AIM)[16,17] (see SI and Figure S2 for Pu M$_4$ edge HERFD-XANES spectra of Pu$^{IV}$O$_{2(cr)}$).

The **Pu$^{VI}$** crystallises as a green solid (Figure 1a) in the monoclinic space group *C*2/*c* unit cell of **Pu$^{VI}$** consists of six [Pu$^{VI}$O$_2$(CO$_3$)$_3$]$^{4-}$ coordination polyhedra (Figure 1b), where each central Pu atom is surrounded by two equidistant axial O atoms forming a [O–Pu$^{VI}$–O]$^{2+}$ moiety. Pu is further coordinated by three CO$_3^{2-}$ ligands

symmetrically bound in the equatorial plane. Four K$^+$ ions are located within [-K–O–Pu–O–K-] planes in each polyhedron and serve for charge compensation (Figure 1c). The cell parameters of **Pu$^{VI}$** fit precisely in the series of isostructural **U$^{VI}$** and **Np$^{VI}$** analogues and makes possible the comparison of their crystallographic properties (c.f. Tables 1, S1).[18, 19] The analysis shows that the unit cell parameters *a*, *b* and *c* decrease in the **U$^{VI}$**–**Np$^{VI}$**–**Pu$^{VI}$** series with a constant decrease of cell volume, *V*, and non-linear shortening of the axial bond distance, d(An$^{VI}$–O)$_{ax}$.

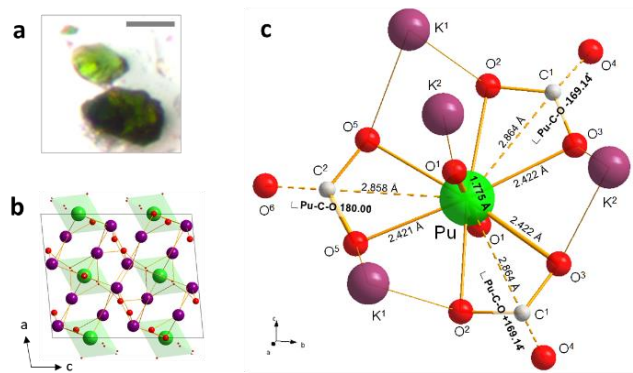

**Figure 1.** Microscopic image of the solid particles selected for SC-XRD analysis (scale bar – 100 μm) (a); representation of the **Pu$^{VI}$** unit cell (b); [Pu$^{VI}$O$_2$(CO$_3$)$_3$]$^{4-}$ viewed from *a*-axis direction with highlighted bond parameters (c).

**Table 1.** Crystallographic parameters of **U$^{VI}$** – K$_4$U$^{VI}$O$_2$(CO$_3$)$_{3(cr)}$,[18] **Np$^{VI}$** – K$_4$Np$^{VI}$O$_2$(CO$_3$)$_{3(cr)}$,[19] and **Pu$^{VI}$** – K$_4$Pu$^{VI}$O$_2$(CO$_3$)$_{3(cr)}$ (values in parentheses shows deviation in the last digit. Detailed crystallographic data for **Pu$^{VI}$** is provided in Table S1).

| Entry | *a*, Å | *b*, Å | *c*, Å | *V*, Å$^3$ | d(An$^{VI}$–O)$_{ax}$, Å |
|---|---|---|---|---|---|
| **U$^{VI}$** | 10.247 | 9.202 | 12.226 | 1152.825 | 1.796(6) |
| **Np$^{VI}$** | 10.214 | 9.177 | 12.182 | 1141.867 | 1.776(4) |
| **Pu$^{VI}$** | 10.113 | 9.158 | 12.153 | 1127.883 | 1.775(5) |

The shortening of an axial bond, d(An$^{VI}$–O)$_{ax}$, in the **U$^{VI}$**–**Np$^{VI}$**–**VI** series is the result of presence of the additional 5f electrons, which increase the electronic density on An atom, 5f$^0$(U$^{VI}$) → 5f$^1$(Np$^{VI}$) → 5f$^2$(Pu$^{VI}$). The increasing electronic density is normally observed in An atom contraction, e.g. decrease of size along the studied series. In the K$_4$An$^{VI}$O$_2$(CO$_3$)$_{3(cr)}$ series, the d(An$^{VI}$–O)$_{ax}$ decreases by 2 pm from **U$^{VI}$** to **Np$^{VI}$**: significantly more than from **Np$^{VI}$** to **Pu$^{VI}$**, which have similar d(An$^{VI}$–O)$_{ax}$ values: 1.776(4) Å and 1.775(5) Å, respectively. Accordingly, the '-yl' oxygens of the aqueous [Pu$^{VI}$O$_2$(CO$_3$)$_3$]$^{4-}$$_{(aq)}$ are 2 pm shorter compared to **Pu$^{VI}$** (d(Pu$^{VI}$–O)$_{ax}$ = 1.75(1) Å vs. 1.775(5) Å), whereas equatorial bond, d(Pu$^{VI}$–O)$_{eq}$, is 2 pm longer, 2.44(1) Å versus averaged 2.426(8) Å.[20] The detailed crystallographic characterization of **Pu$^{VI}$** makes possible the comparison of d(An$^{VI}$–O)$_{ax}$ for solids, K$_4$An$^{VI}$O$_2$(CO$_3$)$_{3(cr)}$, in the **U$^{VI}$**–**Np$^{VI}$**–**U$^{VI}$** series. Moreover, the comparison with corresponding dataset for aqueous, [An$^{VI}$O$_2$(CO$_3$)$_3$]$^{4-}$$_{(aq)}$, species relative to Shannon ionic radii of the corresponding An$^{VI}$ ion can be done. The results shows different trend in d(An$^{VI}$–O)$_{ax}$: co-linear decrease for **U$^{VI}$**–**Np$^{VI}$** and, and clearly different effect for **Np$^{VI}$**–**Pu$^{VI}$**, considering error bars (Figure 2).[18-23] While d(Pu$^{VI}$–O)$_{ax}$ in [Pu$^{VI}$O$_2$(CO$_3$)$_3$]$^{4-}$$_{(aq)}$ decreases by 4 pm compared to [Np$^{VI}$O$_2$(CO$_3$)$_3$]$^{4-}$$_{(aq)}$, in **Pu$^{VI}$** it is similar (within ±0.005 Å) to **Np$^{VI}$** analogue. A negligible change in d(An$^{VI}$–O)$_{ax}$ for **Pu$^{VI}$** can be attributed here to intra- and intermolecular interactions differing dependent on the physical state of the compound, e.g. solid vs. aqueous species. Interestingly, in **Pu$^{VI}$** two of three (Pu$^{VI}$–C–O) angles (169.14°) are significantly bent from the uniplanar (180°) values compared to **U$^{VI}$** with nearly uniplanar (U$^{VI}$–C–O) = 176.00°.[18] A larger (An$^{VI}$–C–O) bending angle might be a result of the crystal cell shrinking in **Pu$^{VI}$** due to actinide contraction effect.

The **Pu$^{VI}$** is further characterized using the Pu M$_4$ edge HERFD-XANES spectroscopy. The method is a powerful tool for analysis of the local geometry around An atom and for probing the valence state, ground state configuration and bonding characteristics of actinide compounds especially when combined with electronic structure calculations.[24]

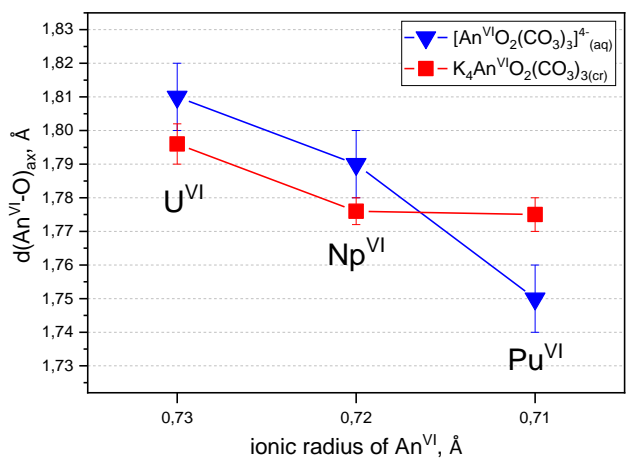

**Figure 2.** Comparison of d(An$^{VI}$–O)$_{ax}$ for [An$^{VI}$O$_2$(CO$_3$)$_3$]$^{4-}$$_{(aq)}$ in ≤ 2M Na$_2$CO$_3$ and[20, 22, 23] K$_4$An$^{VI}$O$_2$(CO$_3$)$_{3(cr)}$ species[18, 19] for An$^{VI}$ = U$^{VI}$, Np$^{VI}$, Pu$^{VI}$ relative to Shannon ionic radius of the corresponding An$^{VI}$ ion.[21]

Figure 3 shows HERFD-XANES spectrum of the **Pu$^{VI}$** assisted with AIM simulations and compared to the spectra of **Pu$^{IV}$** – Pu$^{IV}$O$_{2(cr)}$ and **Pu$^{V}$** – KPu$^{V}$O$_2$CO$_{3(cr)}$ compounds.[15] The HERFD-XANES spectrum for **Pu$^{VI}$** is, to the best of our knowledge, reported for the first time while data for **Pu$^{V}$** and **Pu$^{IV}$** are reproduced from the recent publication to tackle the trends in electronic structure for the **Pu$^{IV}$**–**Pu$^{V}$**–**Pu$^{VI}$** series.[15] Each spectrum shown on Figure 3 exhibits distinct features characteristic to the valence state of plutonium and the crystal structure of plutonium compound: cubic (Fm-3m) for **Pu$^{IV}$**, hexagonal (P6$_3$/mmc) for **Pu$^{V}$**, and monoclinic (C2/c) for **Pu$^{VI}$**. The most intense spectral resonance, white line (WL), of each spectrum shifts to higher energies from ~3969.0 eV for **Pu$^{IV}$** to ~3969.4 eV for **Pu$^{V}$** (ΔE ~ +0.4 eV) and increases further to ~3969.9 eV for **Pu$^{VI}$** (ΔE ~ +0.5 eV). It should be noted that any changes in the local environment near An atom might produce an energy shift in the order of ±0.2 eV, as has been previously observed for several uranium systems.[17, 24] In the same manner the HERFD-XANES spectra of solid and aqueous plutonium compounds will exhibit different positions and shapes of the spectral features. For this reason, the comparison between solid

and aqueous species should be done with special care, ideally supported by theoretical calculations.[25]

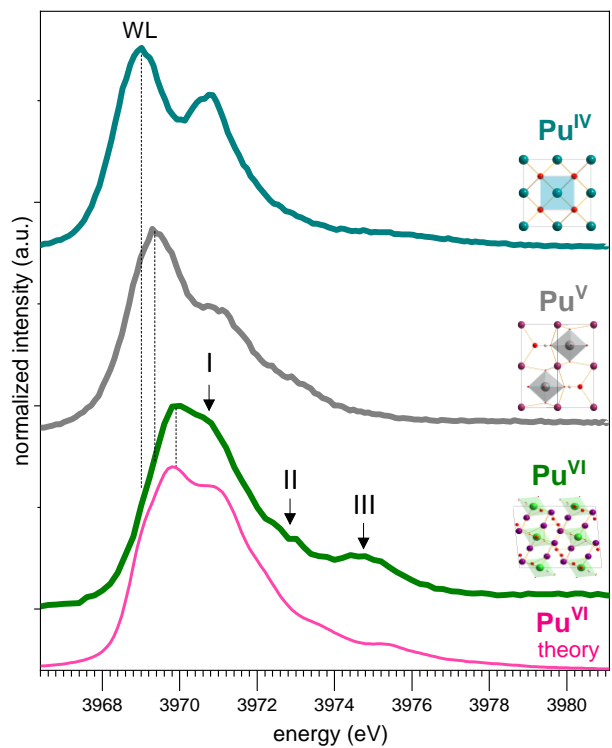

**Figure 3.** Experimental Pu $M_4$ edge HERFD-XANES spectrum of **Pu$^{VI}$** – $K_4Pu^{VI}O_2(CO_3)_{3(cr)}$, recorded at the Pu $M_β$ emission line (~3534 eV) and compared to theoretically calculated HERFD spectrum using the AIM approximation. HERFD-XANES data for reference compounds: **Pu$^{IV}$** – $Pu^{IV}O_{2(cr)}$ and **Pu$^{V}$** – $KPu^{V}O_2CO_{3(cr)}$ is reproduced from Kvashnina et al.[15] and shown here for clarity (unit cell for each compound is highlighted).

For analysis of the electronic structure of **Pu$^{VI}$** we marked the most intense features in the HERFD-XANES spectrum as **I**, **II** and **III** (Figure 3). Based on AIM calculations the Pu $M_4$ edge HERFD spectral features of **Pu$^{VI}$** are referred to multiple factors, such as the strength of the intra-atomic and crystal-field interactions, and the degree of the An–ligand hybridization in the ground and final states of the spectroscopic process. The spectrum of **Pu$^{VI}$** was calculated in a manner described for the U$^{IV}$ system ($U^{IV}O_{2(cr)}$)[17,26] taking into account the Pu 5f hybridization with the ligand valence states and the full multiplet structure due to intra-atomic and crystal field interactions. The Slater integrals $F^k(5f,5f)$, $F^k(3d,5f)$ and $G^k(3d,5f)$ calculated for the **Pu$^{VI}$** ion were scaled down to 80% of their ab-initio Hartree-Fock values. There is a certain consensus to apply such a level of the Slater integral reduction for compounds.[27] The **Pu$^{VI}$** environment was approximated by $D_{2h}$ symmetry for simplicity. Wybourne's crystal field parameters for the 5f shell were set to $B^2_0$ = -0.929 eV, $B^2_2$ = 1.070 eV, $B^4_0$ = 0.416 eV, $B^4_2$ = -0.445 eV, $B^4_4$ = -0.595 eV, $B^6_0$ = -0.043 eV, $B^6_2$ = 0.124 eV, $B^6_4$ = 0.086 eV and $B^6_6$ = -0.136 eV. The values of these crystal-field parameters were estimated using the SIMPRE 1.1 program[28] and Dirac-Fock $<r^n>$ values for the **Pu$^{VI}$** ion adopted from elsewhere.[29]

The ground (final) state of the spectroscopic process was described by a linear combination of the $4f^2$ and $4f^3\underline{v}^1$ ($3d^94f^3$ and $3d^94f^4\underline{v}^1$) configurations where $\underline{v}$ stands for an electronic hole in the ligand valence level. The values for the AIM parameters were as following: the energy for the electron transfer from the valence band to the unoccupied Pu 5f level Δ = 2.0 eV; the 5f-5f Coulomb interaction $U_{ff}$ = 4.0 eV; the 3d core hole potential acting on the 5f electron $U_{fc}$ = 6.5 eV and the Pu 5f – ligand states hybridization term (strength) V = 1.0 eV (0.7 eV) in the ground (final) state of the spectroscopic process. The choice of values of the AIM parameters in our calculations is based on results of the study for a series of An$^{VI}$ compounds.[30] The V value is taken similar to that used in calculations for the uranyl system.[24] A justification for the reduction of the V value in the final state of the spectroscopic process with a core hole present have been discussed earlier for f-electron systems.[31]

According to the AIM calculations, features **I** and **II** in the spectrum of **Pu$^{VI}$** are mainly a consequence of atomic-multiplet and crystal-field effects. Feature **II** is found to depend on the Pu–O distance in the $[O–Pu^{VI}–O]^{2+}$, i.e. on the crystal-field strength, while feature **III** is due to charge-transfer between Pu and ligand sites both from $[Pu^{VI}O_2]^{2+}$ and $CO_3^{2-}$ ligands as a result of the Pu 5f - ligand 2p hybridization. When compared to another 5f$^2$ system, such as $U^{IV}O_{2(cr)}$, the Pu 5f - ligand 2p mixing is found to be stronger in **Pu$^{VI}$** resulting in the 5f occupancy equal to $n_f$ = 2.42 versus $n_f$ = 2.26 for $U^{IV}O_{2(cr)}$.[32] As an outcome of our AIM calculation, the calculated energies for the spread of the multiplet states in the ground state configuration were found to be in fair agreement with those of states obtained in ab-initio quantum-chemical calculations for the $[PuO_2(CO_3)_3]^{4-}$ complex.[33]

Herein, we report synthesis and crystallographic characterisation of the technologically and environmentally important potassium plutonyl carbonate complex - $K_4Pu^{VI}O_2(CO_3)_{3(cr)}$. The HERFD-XANES technique was applied to verify oxidation state of synthesized compound and was used as a basis for the theoretical calculations using the AIM approximation. The synthesis of $K_4Pu^{VI}O_2(CO_3)_{3(cr)}$ filled the isostructural U$^{VI}$–Np$^{VI}$–U$^{VI}$ series and made possible the analysis of the structural trends for both solid and aqueous species. It is shown on the available datasets for solid, $K_4An^{VI}O_2(CO_3)_{3(cr)}$, and aqueous, $[An^{VI}O_2(CO_3)_3]^{4-}_{(aq)}$ species for U$^{VI}$, Np$^{VI}$ and Pu$^{VI}$ that $d(An^{VI}–O)_{ax}$ may differ significantly depending on the physical state of compound and the electronic structure of An atom, i.e. configuration of 5f electrons followed by the change of structural parameters. This behaviour is revealed for $K_4Pu^{VI}O_2(CO_3)_{3(cr)}$ and $[Pu^{VI}O_2(CO_3)_3]^{4-}_{(aq)}$ when compared to U$^{VI}$ and Np$^{VI}$ analogues from nearly co-linear trend for U$^{VI}$ and Np$^{VI}$ and opposite direction for U$^{VI}$ and Pu$^{VI}$ species.

## Associated Content

Supporting Information includes details for **Pu$^{VI}$** synthesis, design of the in situ spectroelectrochemical setup, SC-XRD characterization, crystallographic parameters of **Pu$^{VI}$**, details for HERFD-XANES measurements and AIM theoretical calculations.

## Acknowledgements

This research was funded by European Commission Council under ERC grant 759696. S.M.B. acknowledges support from the Swedish Research Council (research grant 2017-06465). We thank SOLEIL for the granted beamtime. I.P. thanks Dr. Vanessa Montoya (Helmholtz Centre for Environmental Research - UFZ, Leipzig) for helpful discussion.

**Keywords:** plutonium carbonate • crystal structure • SC-XRD • electronic structure • Pu M$_4$ edge HERFD

**Entry for the Table of Contents**

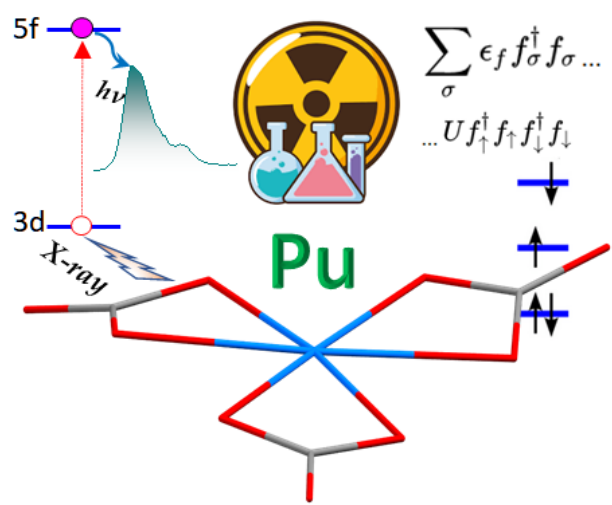

Environmentally relevant potassium plutonium carbonate complex - $K_4Pu^{VI}O_2(CO_3)_{3(cr)}$ - has been synthesized and characterized by the single-crystal X-ray diffraction and high energy resolution fluorescence detected (HERFD) X-ray absorption spectroscopy at the Pu $M_4$ edge together with electronic structure calculations, based on Anderson impurity model.